\documentclass{pinchcr}
\usepackage{amsmath,amssymb,makeidx}
\newcommand*{\be}{\begin{equation}}
\newcommand*{\ee}{\end{equation}}

\newcommand{\om}{\omega}
\newcommand*{\omc}{\omega_{\text{c}}}
\newcommand*{\omq}{\omega_{\text{q}}}
\newcommand*{\omd}{\omega_{\text{d}}}
\newcommand*{\bigO}{\mathcal{O}}
\newcommand*{\Nbare}{N_\text{bare}}
\newcommand*{\Ncrit}{N_\text{crit}}
\newcommand*{\Nsc}{N_\text{sc}}
\newcommand{\la}{\langle}
\newcommand{\ra}{\rangle}

\newcommand{\ad}{a^{\dagger}}
\newcommand*{\dg}{^{\dag}}

\newcommand{\s}{\sigma}

\newcommand*{\rmi}{\mathrm{i}}

\newcommand*{\rmd}{\mathrm{d}}

\usepackage{hyperref}
\usepackage[all]{hypcap}
\begin{document}
\chapter{Nonlinear oscillators and high fidelity qubit state measurement in circuit quantum electrodynamics\\ {\normalsize Eran Ginossar$^1$, Lev S. Bishop $^2$ and S. M. Girvin$^3$}}
\footnotetext[1]{Advanced Technology Institute and Department of Physics, University of Surrey, Guildford GU2 7XH, United Kingdom}
\footnotetext[2]{Joint Quantum Institute and Condensed Matter Theory Center, Department of Physics, University of Maryland, College Park, Maryland 20742, USA}
\footnotetext[3]{Department of Physics, Yale University, 217 Prospect Street, New Haven, Connecticut 06511}

\section{Introduction: The high power response of the transmon-cavity system}
The Jaynes--Cummings~(JC)\index{Jaynes--Cummings} Hamiltonian provides a quantum model for a two-level system (qubit) interacting with a quantized electromagnetic mode. It is widely applicable to experiments with natural atoms \shortcite{haroche_raimond_exploring,gleyzes_quantum_2007,boca_observation_2004,brennecke_cavity_2007,maunz_normal-mode_2005}
as well as for solid-state `artificial atoms'~\shortcite{reithmaier_strong_2004,yoshie_vacuum_2004,wallra_strong_2004}. The discussion in this chapter applies to any realization of the model that can reach the appropriate parameter regimes and be driven sufficiently strongly\index{strong!driving}, but for concreteness we adopt the language and focus on typical parameters from the field of circuit quantum electrodynamics (circuit QED),\index{Circuit QED} where the relevant parameter range is easily achieved in experiments.

We write the Jaynes-Cummings model \index{Jaynes--Cummings} with drive and dissipation ($\hbar=1$)
\be H=\om_c \ad a + \frac{\om_q}{2} \s_z + g(\s_+ a + \ad \s_- ) + \frac{\xi}{\sqrt{2}} (a+\ad) + H_{\gamma} + H_{\kappa} \ee
with cavity frequency $\omc/2\pi$, qubit frequency $\omq/2\pi$ and where $\xi(t)$ is the time-dependent drive of the cavity, $g$ is the cavity-qubit coupling, and $H_{\gamma,\kappa}$ represent the coupling to the qubit and cavity baths, respectively.

The JC Hamiltonian can be diagonalized analytically, but in the presence of a drive $\xi(t)$ and dissipation the open-system model becomes non-trivial, with the effective behavior depending strongly on the specific parameter regime. Several central parameters are involved in the classification of theses different regimes. The case where the cavity relaxation rate $\kappa$ is much larger (smaller) than the two-level dissipation and dephasing rates $\gamma$, $\gamma_\phi$ is known as the bad (good) cavity limit. The \emph{strong dispersive} regime~\shortcite{gambetta_qubit-photon_2006,schuster_resolving_2007}\index{dispersive regime} of the JC model describes the situation that the presence of the qubit causes the cavity frequency to be shifted by an amount $\chi$ much greater than the cavity linewidth. The shift  depends on the number of excitations in the systems $\chi=\chi(N)$ and for low photon numbers is similar to a Duffing oscillator nonlinearity \index{Duffing oscillator}.

In cavity QED, various forms of ``single atom bistability'' \index{bistability} in this model are known: single atom absorptive bistability~\shortcite{savage,mabuchi-bistability-2011}, exists in the weak coupling regime in the good cavity limit \index{good cavity limit}($g\ll\gamma\ll\kappa$); also  closely related is the single atom phase bistability of spontaneous dressed state polarization~\shortcite{alsing_spontaneous_1991,kilin_single-atom_1991} which concerns the case where the atom and the cavity are in resonance $\delta=\om_c-\om_q=0$.

In this chapter we analyze the high excitation nonlinear response of the Jaynes--Cummings \index{Jaynes--Cummings}model in quantum optics when the qubit and cavity are strongly coupled\index{strong!coupling}. We focus on the parameter ranges appropriate for transmon qubits in the circuit quantum electrodynamics architecture\index{Circuit QED}, where the system behaves essentially as a nonlinear quantum oscillator and we analyze the quantum and semi-classical dynamics. One of the central motivations is that under strong excitation tones\index{strong!driving}, the nonlinear response can lead to qubit quantum state discrimination and below we present initial results for the cases when the qubit and cavity are on resonance or far off-resonance (dispersive).\index{dispersive regime}

\section{Implications of the nonlinearity at the high excitation regime}\label{gino:implications}
\sectionmark{Implications of the nonlinearity}

A characteristic feature of the JC model is that for very high excitation number $N\gg 1$, the excitation number-dependent frequency shift obeys $\chi(N) \rightarrow 0$: the transition frequency returns to the bare cavity frequency.  In the presence of dissipation this happens effectively when $\chi(N) \lesssim \kappa$, and for all larger $N$ the response of the system is linear with respect to the drive. We describe this behavior as setting in at an excitation number $\Nbare$, with the definition $\chi(\Nbare)=\kappa$. In the strong dispersive regime \index{dispersive regime} we have $\Nbare\gg\Ncrit$, where $\Ncrit$ as usual denotes the excitation level where the dispersive approximation \index{dispersive regime}breaks down (defined below). The latter inequality has an important consequence for the theory: a perturbative expansion in the small parameter $N/\Ncrit$, typically useful~\shortcite{boissonneault_nonlinear_2008,boissonneault_dispersive_2009} in the dispersive regime, is not applicable for the interesting regime $N>\Nbare$ where the system regains the linear response.

\begin{figure}[tb]
\centering
\includegraphics[scale=0.5]{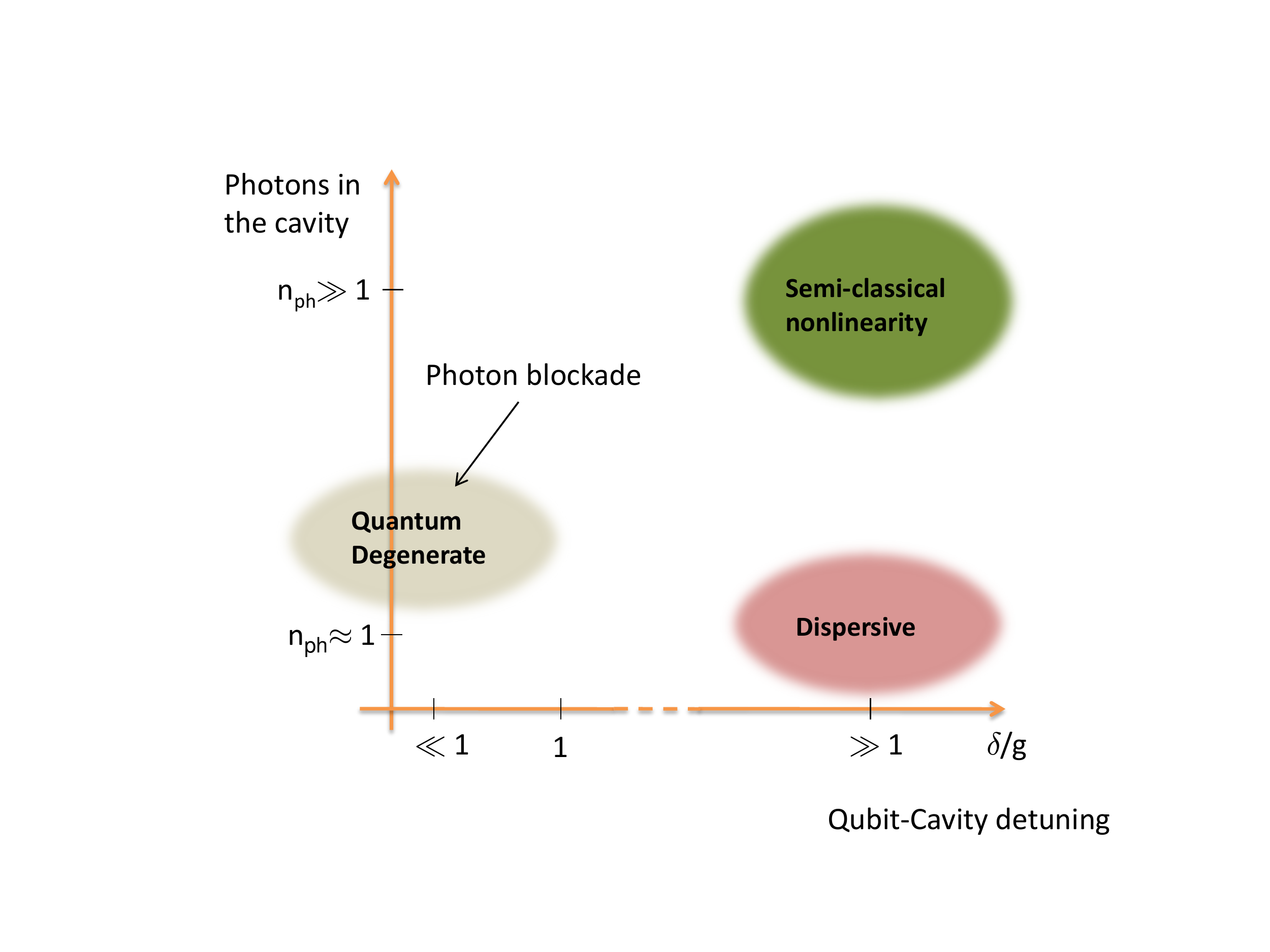}\centering
\caption{Qubit state measurement in circuit QED \index{Circuit QED}can operate in different parameter regimes and relies on different dynamical phenomena of the strongly coupled transmon-resonator system\index{strong!coupling}. The dispersive readout is the least disruptive to the qubit state and it is realized where the cavity and qubit are strongly detuned. The high power readout operates in a regime where the system response can be described using a semi-classical model and yields an relatively high fidelity \index{fidelity}with simple measurement protocol. When the cavity and qubit are on resonance (the quantum degenerate regime\index{quantum degenerate regime}) it is theoretically predicted that the photon blockade \index{photon blockade} can also be used to realize a high fidelity readout.
\label{gino:fig:context}}
\end{figure}
We discuss two different parameter regimes, the semi-classical and quantum degenerate, shown schematically in fig.~\ref{gino:fig:context}. We note that a very related semi-classical nonlinearity can be introduced via adding a Josephson junction directly to the resonator. This approach leads to the Josephson Bifurcation Amplifier (JBA)\index{bifurcation}, the Cavity Bifurcation Amplifier (CBA) which realize an efficient qubit readout based on the dynamical bistability \index{bistability} of the resonator state, and were the first to make use of bright states for qubit measurement~\shortcite{siddiqi_dispersive_2006,boulant_quantum_2007,metcalfe_cba_2007}. In this review we confine ourselves to the minimal model of only one qubit interacting with one linear cavity which has a different structure of anharmonicity with unique effects.

In Section~\ref{gino:sec:transient}, we consider the model in the bad cavity limit \index{bad cavity limit}and on timescales short compared to the atomic coherence time where the dynamics are those of a nonlinear oscillator, which we study by both semiclassical methods \index{semiclassical} and quantum trajectory simulations\index{quantum trajectories}. Our main result is that there exists a threshold drive $\xi_\text{C2}$ at which the photon occupation increases by several orders of magnitude over a small range of the drive amplitude. We perform both quantum trajectory simulations and a non-perturbative semiclassical analysis, including the drive and the cavity damping. Our results are in qualitative agreement with recent experiments~\shortcite{reedout_expt} for a circuit quantum electrodynamics~(QED) \index{Circuit QED}device~\shortcite{leo_ghz_2010} containing 4 transmon~\shortcite{koch_charge-insensitive_2007,schreier_suppressing_2008} qubits\index{transmon}, demonstrating that the JC model captures the essential physics despite making an enormous simplification of the full system Hamiltonian.

In Section~\ref{gino:sec:degenerate} the situation where the qubit and the cavity are on resonance is shown to lead to a coexistence of photon blockaded \index{photon blockade}states and highly excited quasi-coherent states (QCS) \index{quasi-coherent states}with the same driving tone. This is also a result of the nonlinearity that arises in the Jaynes-Cummings model\index{Jaynes--Cummings}, but in contrast this regime cannot be fully described with a semiclassical theory \index{semiclassical}and requires exact quantum simulations \index{quantum trajectories} to reveal.

\subsection{Transient response in the dispersive regime}\label{gino:sec:transient}

The behavior of the JC nonlinearity goes beyond the Kerr nonlinearity \index{Kerr nonlinearity}that is often considered. Dispersive bistability~\shortcite{marburger_theory_1978} \index{bistability} from a Kerr nonlinearity is well-known in atomic cavity QED~\shortcite{gibbs_differential_1976}. It has been implemented in the solid state via the nonlinearity of a Josephson junction~\shortcite{siddiqi_rf-driven_2004}, and in the circuit QED architecture\index{Circuit QED} has produced high-fidelity \index{fidelity}readout of qubits~\shortcite{siddiqi_dispersive_2006,boulant_quantum_2007,mallet_single-shot_2009}. Similar schemes have been implemented with nonlinear micromechanical resonators~\shortcite{almog_high_2006}. However, unlike the Kerr anharmonicity, the JC anharmonicity does not remain constant but rather diminishes toward zero as the cavity occupation is increased. As a result, for sufficiently strong drive \index{strong!driving}the response of the JC model must return to the linear response of the bare cavity. Instead of coherent driving, an alternative way to saturate the qubit and cause the JC system response to return to the bare cavity response is to couple the system to a bath at elevated temperature, as has been investigated theoretically~\shortcite{Dykman_Ivanov_1976,rau_cavity_2004} and experimentally~\shortcite{fink_quantum-to-classical_2010}.

Recent experiments~\shortcite{reedout_expt}, which operate in both the strong dispersive regime \index{dispersive regime}and the bad cavity limit\index{bad cavity limit}, show a nontrivial response under conditions of strong drive, arousing interest due to its usefulness for high-fidelity qubit readout.

\begin{figure}[tpb]
\centering
\includegraphics[scale=1.4]{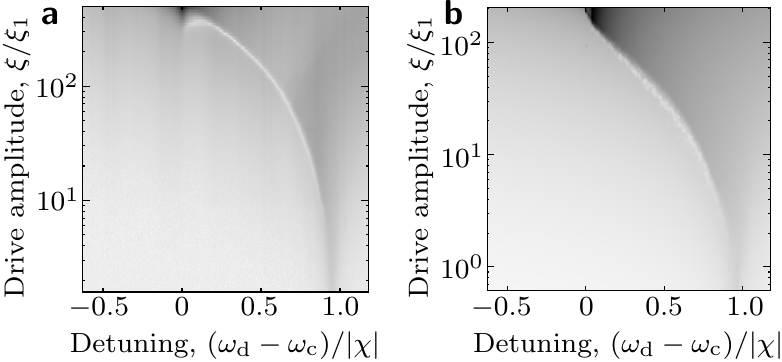}
\caption[XXX broken OUP style needs this XXX]{Transmitted heterodyne amplitude $\lvert\langle a\rangle\rvert$ as a function of drive detuning (normalized by the dispersive shift \index{dispersive regime}$\chi=g^2/\delta$) and drive amplitude (normalized by the amplitude to put $n=1$ photon in the cavity in linear response, $\xi_1=\kappa/\sqrt{2}$). Dark colors indicate larger amplitudes.
(a)~Experimental data~\shortcite{matt-pc}, for a device with cavity at $9.07\,\text{GHz}$ and 4 transmon qubits \index{transmon}at $7.0, 7.5, 8.0, 12.3\,\text{GHz}$. All qubits are initialized in their ground state, and the signal is integrated for the first $400\,\text{ns}\simeq4/\kappa$ after switching on the drive.
(b)~Numerical results for the JC model of eqn~\ref{gino:eq:master}, with qubit fixed to the ground state and effective parameters $\delta/2\pi=-1.0\,\text{GHz}$, $g/2\pi=0.2\,\text{GHz}$, $\kappa/2\pi=0.001\,\text{GHz}$. These are only intended as representative numbers for circuit QED \index{Circuit QED}and were not optimized against the data of panel~(a). Hilbert space is truncated at 10,000 excitations (truncation artifacts are visible for the strongest drive), and results are shown for time $t=2.5/\kappa$.}\label{gino:fig:latch000}
\end{figure}

For driving at a single frequency, we write the driven JC Hamiltonian
\begin{equation*}
  H=\omc a^{\dag}a+\frac{\omq}{2} \sigma_z + g(a \sigma_+ + a^\dag \sigma_-)
  +\frac{\xi}{\sqrt{2}}(a+a^\dag)\cos(\omd t),
\end{equation*}
given with the drive frequency $\omd$. Operating in the strong-dispersive bad-cavity regime defines a hierarchy of scales
\be\label{gino:eq:regime}
    \gamma,\gamma_\phi\ll\kappa\ll g^2/\delta\ll g\ll \delta\ll\omc ,
\ee
where $\delta=\omq-\omc$ is the qubit-cavity detuning.
We can make the standard transformation~\shortcite{canonicaldressing} $\tilde{H}=T^{-1} H T$ to decouple the qubit and cavity
\be\label{gino:eq:H_car}
  \tilde{H}=
  \omc a^\dag a+ (\omc-\Delta)\frac{\sigma_z}{2}\\
+\frac{\xi}{\sqrt{2}}(a+a^{\dag})\cos(\omd t) ,
\ee
dropping terms from the transformed drive that are suppressed as $\bigO(N^{-1/2})$ and $\bigO(g/\delta)$. The resulting Hamiltonian would be trivial were it not for the fact that the transformation $T$, defined by
\begin{align}
  T&=\exp[-\theta(4N)^{-1/2}(a \sigma_+ + a^\dag \sigma_-)] , \\
  \sin\theta&=-2g N^{1/2}/\Delta,\quad \cos\theta=\delta/\Delta , \\
  \Delta&=(\delta^2+4g^2 N)^{1/2} \label{gino:eq:Delta},
\end{align}
depends on the total number of excitations,  $N=a\dg a +\sigma_z/2 + 1/2$.
For photon decay at rate $\kappa$ we can write the decoupled quantum master equation after dropping small terms,
\be\label{gino:eq:master}
    \dot{\rho}=-\rmi[\tilde{H},\rho]+\kappa([a\rho,a\dg]+[a,\rho a\dg])/2,
\ee
which we integrate numerically in a truncated Hilbert space using the method of quantum trajectories\index{quantum trajectories}, after making the rotating wave approximation~(RWA) with respect to the drive. The experiments we wish to describe are performed on a timescale short compared to the qubit decoherence times $\gamma^{-1}$, $\gamma_\phi^{-1}$ and we therefore treat $\sigma_z$ as a constant of motion. The remaining degree of freedom constitutes a Jaynes--Cummings oscillator\index{Jaynes--Cummings}. Note that the qubit relaxation and dephasing terms that we have dropped involve the $\sigma_\pm$ and $\sigma_z$ operators and would transform in a nontrivial way under the decoupling transformation $T$~\shortcite{boissonneault_nonlinear_2008}, \shortcite{boissonneault_dispersive_2009}. The results of the numerical integration for $\sigma_z=-1$ are compared with recent experimental data~\shortcite{matt-pc} in Fig.~\ref{gino:fig:latch000}, where we show the average heterodyne amplitude $\lvert\langle a \rangle\rvert$ as a function of drive frequency and amplitude. Despite the presence of 4 qubits in the device, the fact that extensions beyond a two-level model would seem necessary since higher levels of the transmons \index{transmon}are certainly occupied for such strong driving\footnote{Simulations show approximately 10 transmon levels are required to simulate the experiment quantitatively.}\index{strong!driving}, and despite the fact that the Rabi Hamiltonian might seem more appropriate for such large photon occupation, $\sqrt{N}g\sim\omc$, nevertheless the JC model qualitatively reproduces the features of the experiment\footnote{We emphasize that the effective parameters in the simulation are of the same magnitude as in the experiments but we do not expect any quantitative correspondence.}. In particular, for weak driving we see a response as expected at the dispersively shifted cavity frequency $\omc-\chi$, with $\chi=g^2/\delta$, which shifts towards lower frequencies as the drive increases. For stronger driving a dip appears in the response, which we interpret as a consequence of plotting the absolute value of the ensemble-averaged amplitude $\la a\ra$ in the classically bistable region, as we discuss below. For increasing drive the dip shifts to lower frequencies; finally for the strongest driving, the response becomes centered at the bare cavity frequency $\omc/2\pi$ and is single-peaked and extremely strong. We note that both the experiment and numerical integration are terminated at a transient time of only a few cavity lifetimes, and we have checked that the numerical response is substantially different for the steady state (see below).

When there is a large number of photons in the system such that the anharmonicity is greatly diminished, it is possible to use a semiclassical model, similar to~\shortcite{alsing_spontaneous_1991}, \shortcite{kilin_single-atom_1991}, \shortcite{peano_dynamical_2010}, to characterize the transmission. In fact this is also a good approximation for the response at low powers in the dispersive regime\index{dispersive regime} when the ratio of the anharmonicity of the dispersive Hamiltonian to the decay rate (width of the levels) is such that the $N-1\leftrightarrow N$ photon peak overlaps well with the $N\leftrightarrow N+1$ photon peak. By expanding~\ref{gino:eq:H_car} to second order in $g^2N/\delta^2$ this condition can be seen to be $N\gg\Nsc$, where $\Nsc=g^4/\kappa\delta^3$ (for the parameters of Fig.~\ref{gino:fig:latch000}b, $\Nsc=1.6$). In the opposite limit we will see photon blockade \index{photon blockade}and associated effects, as in~\shortcite{bishop_nonlinear_2009}. Recently it was shown that it is possible to have a coexistence of both the semiclassical\index{semiclassical} and quantum solutions for a certain range of parameters of the system and drive~\shortcite{blammo} as we discuss below in Section~\ref{gino:sec:degenerate}. The semiclassical model will remain valid for $N>\Ncrit$, where a perturbative expansion of the Hamiltonian \ref{gino:eq:H_car} in terms of $N/\Ncrit$ fails to converge, where $\Ncrit=\delta^2/4g^2$. We rewrite the Hamiltonian eqn~\ref{gino:eq:H_car} using canonical variables $X=\sqrt{1/2}(a\dg+a)$ and $P=\rmi\sqrt{1/2}(a\dg-a)$,
\be\begin{split}
  \tilde{H}&=\frac{\omc}{2}(X^2+P^2+\sigma_z)+\xi X \cos(\omd t)\\
  &-\frac{\sigma_z}{2}\sqrt{2g^2(X^2+P^2+\sigma_z)+\delta^2}.
\end{split}\ee
The semiclassical approximation consists of treating $X$ and $P$ as numbers. Instead of directly solving Eq.~\ref{gino:eq:master} we write the equations of motion for $X,P$ from the diagonalized Hamiltonian $\tilde{H}$ of the closed system and the effect of cavity relaxation is incorporated through a phenomenological damping term proportional to $\kappa$. We solve for the steady state, treating $X^2+P^2$ as a constant (thus we ignore harmonic generation), giving a nonlinear equation for the amplitude $A=\sqrt{X^2+P^2}$
\be\label{gino:eq:classic}
    A^2=\frac{\omc^2\xi^2}{\left[\omd^2-(\omc-\chi(A))^2\right]^2+\kappa^2\omd^2}
\ee
with amplitude-dependent frequency shift $\chi(A)$, given by\footnote{Note that $N \approx A^2$ in the semiclassical approximation and in that case $\chi(N)=\chi(A^2)=\chi(A)$.}
\be\label{gino:eq:chi}
   \chi(A)=\sigma_z \frac{g^2}{\sqrt{2g^2(A^2+\sigma_z)+\delta^2}}.
\ee
This reproduces for small driving the usual dispersive shift \index{dispersive regime}$\chi(0)\simeq\pm g^2/\delta$ and for strong driving shows the saturation effect $\lim_{A\to\infty}\chi(A)=0$.
\begin{figure}[tpb]
\centering
\includegraphics[width=189pt]{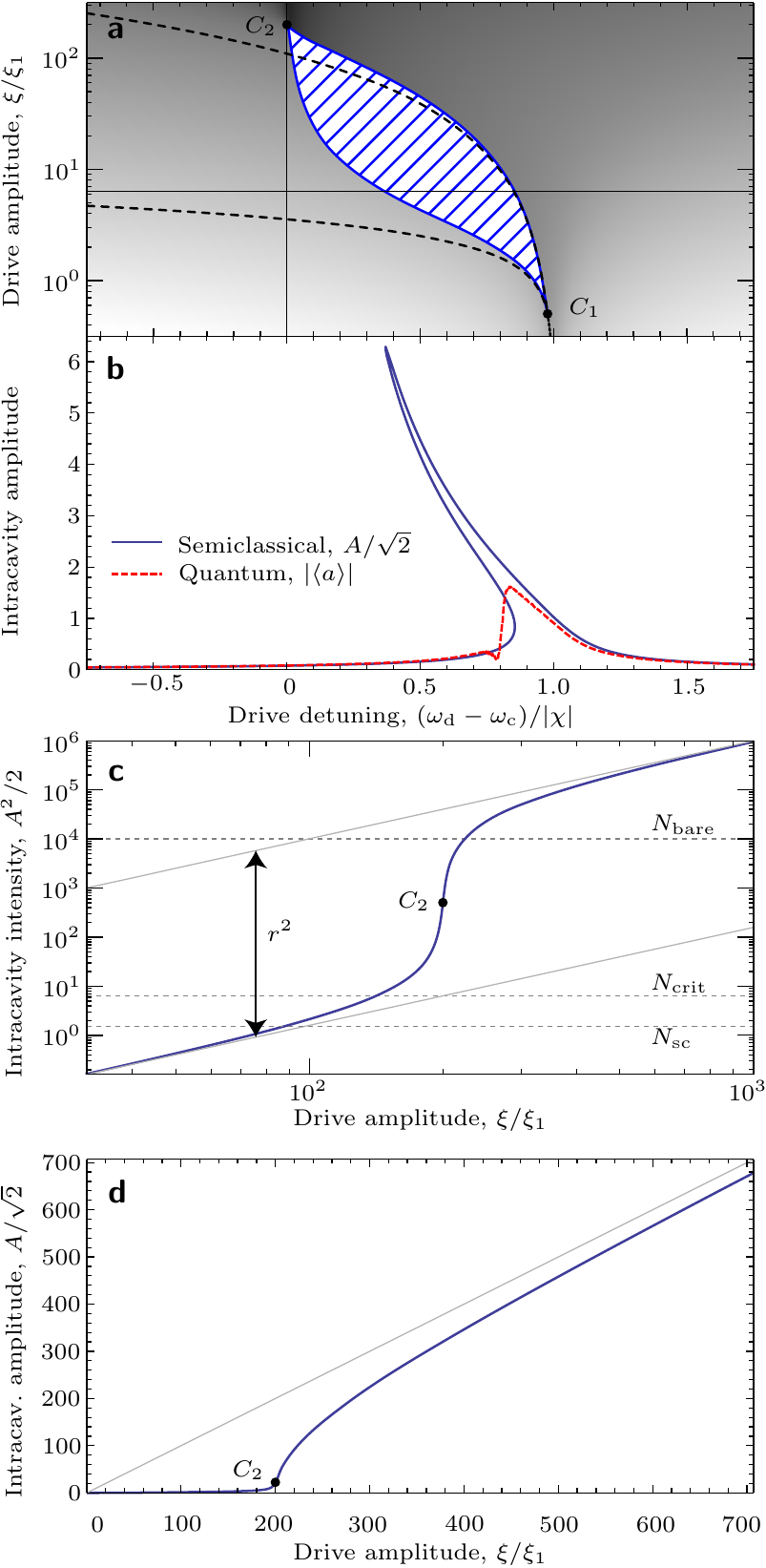}
\caption{Solution to the semiclassical equation~\ref{gino:eq:classic}, using the same parameters as Fig.~\ref{gino:fig:latch000}b.
 (a)~Amplitude response as a function of drive frequency and amplitude. The region of bifurcation \index{bifurcation}is indicated by the shaded area, and has corners at the critical points $C_1$, $C_2$. The dashed lines indicate the boundaries of the bistable region for a Kerr oscillator (Duffing oscillator)\index{Duffing oscillator}, constructed by making the power-series expansion of the Hamiltonian to second order in $N/\Ncrit$. The Kerr bistability \index{bistability}\index{Kerr} region
 matches the JC region in the vicinity of $C_1$ but does not exhibit a second critical point.
 (b)~Cut through (a) for a drive of $6.3\xi_1$, showing the frequency dependence of the classical solutions (solid line). For comparison, the response from the full quantum simulation of Fig.~\ref{gino:fig:latch000}b is also plotted (dashed line) for the same parameters.
 (c)~Cut through (a) for driving at the bare cavity frequency, showing the large gain available close to $C_2$ (the `step'). Faint lines indicate linear response.
 (d)~Same as (c), showing intracavity amplitude on a linear scale.\label{gino:fig:densclass}}
\end{figure}

The solution of eqn~\ref{gino:eq:classic} is plotted in Fig.~\ref{gino:fig:densclass} for the same parameters as in Fig.~\ref{gino:fig:latch000}b. For weak driving the system response approaches the linear response of the dispersively shifted cavity. Above the lower critical amplitude $\xi_{C1}$ the frequency response bifurcates, and the JC oscillator enters a region of bistability\index{bistability}. We denote by $C_1$ the point at which the bifurcation \index{bifurcation}first appears. Dropping terms which are small according to the hierarchy of eqn~\ref{gino:eq:regime}, this point occurs at
$\xi_{C1}=(\delta\kappa)^{3/2}3^{-3/4}g^{-2}$, $\Omega_{C1}=\chi(0)-\sqrt{3}\kappa/2$,
writing the drive detuning as $\Omega=\omega_d-\omega_c$.
The dip in the heterodyne measurement of Fig.~\ref{gino:fig:latch000} appears within the bifurcation\index{bifurcation} region (Fig.~\ref{gino:fig:densclass}a), indicating that this dip is the result of ensemble-averaging of the coherent heterodyne amplitude in the region of classical bistability. In Fig.~\ref{gino:fig:densclass}b we see that the semiclassical\index{semiclassical} and quantum simulation yield the same response outside the region of bistability. Within the region of bistability, quantum noise causes switching~\shortcite{Dykman_Smelyanskiy_1988} between the two semiclassical solutions, one dim and one bright, with almost opposite phases. An analytical derivation of the dip in the steady-state amplitude for a Kerr nonlinearity \index{Kerr nonlinearity} was given in~\shortcite{drummond_walls_1980}. In our case, both the experiment and numerical integration are terminated at a transient time of only a few cavity lifetimes. Therefore the exact form of the averaged response is influenced by the initial conditions. We have checked that the position of the dip in the numerical response is shifted towards lower frequencies in the steady state, consistent with the switching rate being slow compared to the cavity decay rate.

As the drive increases, and unlike the Kerr oscillator\index{Kerr nonlinearity}, the frequency extent of the bistable region shrinks and eventually vanishes at the upper critical amplitude $\xi_{C2}=g/\sqrt{2}$. In the effective theory the upper critical point $C_2$ is located
very close to the bare cavity frequency. This indicates that for driving at the bare cavity frequency, there is no bistability\index{bistability}, but rather a finite region (a `step') in the vicinity of the critical point (Fig.~\ref{gino:fig:densclass}c), where the response becomes strongly sensitive to the drive amplitude. The size of the step can be shown to be a factor of $r=A_\text{bright}/A_\text{dim}=2 g^2/\kappa\delta$ in amplitude, and represents a very high gain ($\rmd A/\rmd \xi=\sqrt{2}g/\kappa^{3/2}\delta^{1/2}$) in the strong-dispersive regime\index{dispersive regime}. Above the step we see that the response approaches the linear response of the bare cavity as $N\simeq\Nbare$.

\subsection{The quantum degenerate regime\label{gino:sec:degenerate}}
\index{quantum degenerate regime}
We now move to exploring what happens when the detuning between the qubit and the cavity is reduced such that the anharmonicity of quantum ladder of states becomes much larger than the corresponding linewidth $\kappa$ (see Fig.~\ref{gino:fig:context}). In order to describe the response of the system to external drive in this regime it is important to take into account the quantum dynamics on the lower anharmonic part of the ladder.
When the system is initialized in the ground state, there is a range of drive strengths for which the system will remain blockaded from excitations out of the ground state. However, since the anharmonicity of the JC ladder decreases with excitation number, the transition frequency for excitations between adjacent levels ultimately approaches the bare cavity frequency. Qualitatively, when the excitation level $n$ is such that the anharmonicity becomes smaller than the linewidth $\kappa$, we expect the state dynamics to be semiclassical, similar to a driven-damped harmonic oscillator \shortcite{alsing_spontaneous_1991,kilin_single-atom_1991}. In this regime we therefore expect the drive to excite states which have an occupation function similar to coherent states but are highly mixed, which we call quasi-coherent states (QCS)\index{quasi-coherent states}. More specifically, in order to support a coherent wave packet centered around level $n$, with a standard deviation of $\sqrt{n}$, the difference of transition frequencies across the wavepacket has to be of the order of the linewidth $\kappa$. This approximate criterion for a minimal $n$ can be written as $\omega_{n+2\s}-\omega_{n-2\s}\approx \kappa$
where $\om_{n\pm 2\s}$ are the ladder transition frequencies, positioned $2\s=2\sqrt{n}$ above and below the mean level $n$.

\begin{figure}[tpb]
\centering
\includegraphics[scale=1.2]{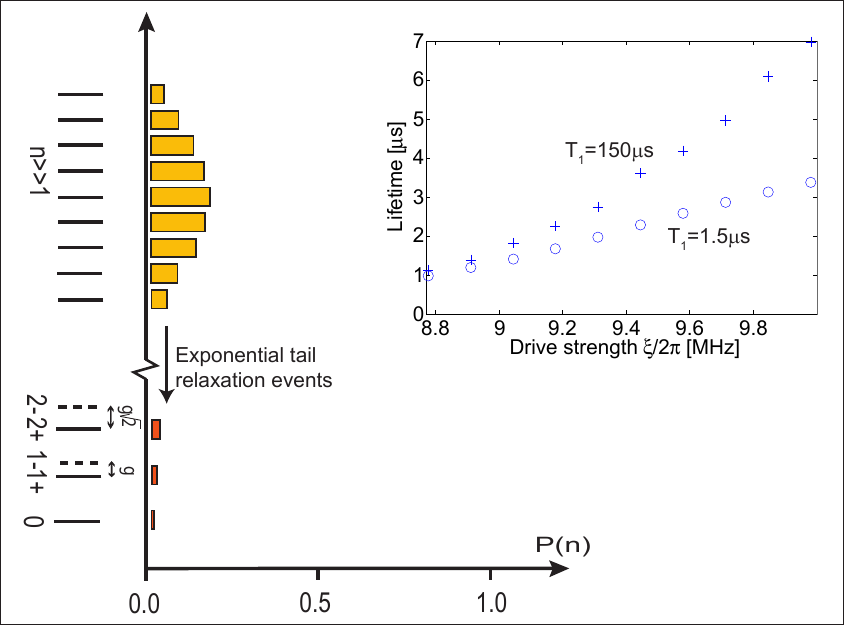}
\caption{\label{gino:lifetime_figure} Schematic figure showing an interpretation of the numerical simulations. Long-lived quasi-coherent states (QCS) decay due to photon emission events which occur in the exponential tail of the wave packet in the lower anharmonic parts of the ladder of states of the Jaynes-Cummings model\index{Jaynes--Cummings}. These emissions are uncompensated by the detuned drive and as a result the wave packet eventually falls into the blockaded regime $\la n\ra \rightarrow 0$. The inset shows the results of quantum simulations: the lifetimes ($\tau_b$) of QCS is increasing with drive strength and mean photon number for drive strengths where the dim state is still photon blockaded demonstrating the coexistence regime. We also see that the qubit decay ($T_1$) has a distinct influence on the lifetime of the QCS (here the coupling was taken to be $g/2\pi=100 \textrm{MHz}$ and the cavity lifetime here is $\kappa^{-1}=60ns$).}
\end{figure}

To study this system, we use the stochastic Schr\"{o}dinger equation \index{stochastic Schr\"{o}dinger equation}which describes the quantum evolution of the qubit-cavity system in the presence of qubit and cavity dissipation and a direct photodetection model~\shortcite{breuer_petroccione_book}
\begin{eqnarray}\label{gino:eq:sse}
   &&d\psi(t)=-i\left(H(t)-\frac{\gamma}{2}\sigma_+\sigma_--\frac{\kappa}{2}\ad a-\right)\psi(t)dt+\\
   && +\left( \frac{\sigma_-\psi(t)}{|\sigma_-\psi(t)|}-\psi(t) \right)dN_1+\left( \frac{a\psi(t)}{|a\psi(t)|}-\psi(t) \right)dN_2
\end{eqnarray}
with the $H(t)$ given by Eqn.~\ref{gino:eq:regime}.
Using numerical integration of this equation we generate an ensembles of trajectories\index{quantum trajectories} of the quantum evolution of the wave function conditioned on the measurement signal. Quantitatively, we find that the lifetime of the QCS is long but finite, and increases with the amplitude of the drive (see inset in Fig.~\ref{gino:lifetime_figure}).

As we explain below, we find that low-lying QCS ($\bar{n}=20$) are the most effective for optimizing the overall readout fidelity\index{fidelity}. Note that the JC ladder consists of two manifolds (originating from the degeneracy of the bare states $|g, n+1\ra$ and $|e, n\ra$) denoted by $(\pm)$, and we will always refer to states occupying one manifold since the drive is off-resonant with respect to the other manifold. Transitions between manifolds contribute to the decay of the QCS to the dim state. Such transitions can be induced by the drive but their rate is smaller by a factor of $\mathcal{O}(n^{-1/2})$ compared to the rate of transitions inside the same manifold. An additional source of inter-manifold transitions are decay ($T_1$) and pure qubit dephasing $(T_{\varphi})$, whose effects in the presence of drive were studied in the context of the dispersive regime \shortcite{boissonneault_dispersive_2009}\index{dispersive regime}. Indeed, as we see in Fig.~\ref{gino:lifetime_figure}, changing $T_1$ has a noticeable effect on the QCS lifetimes. For very large $\bar{n}$, these processes become ineffective for inducing decay of QCS, since then the difference between the manifold excitation frequencies becomes smaller than $\kappa$, and therefore the drive effectively drives both manifolds. For superconducting transmon qubits\index{transmon}\index{superconductivity} $T_1$ is the dominant decay process, and we show its effect on the overall fidelity in Fig.~\ref{gino:hist_figure}.
\begin{figure}
\centering
\includegraphics{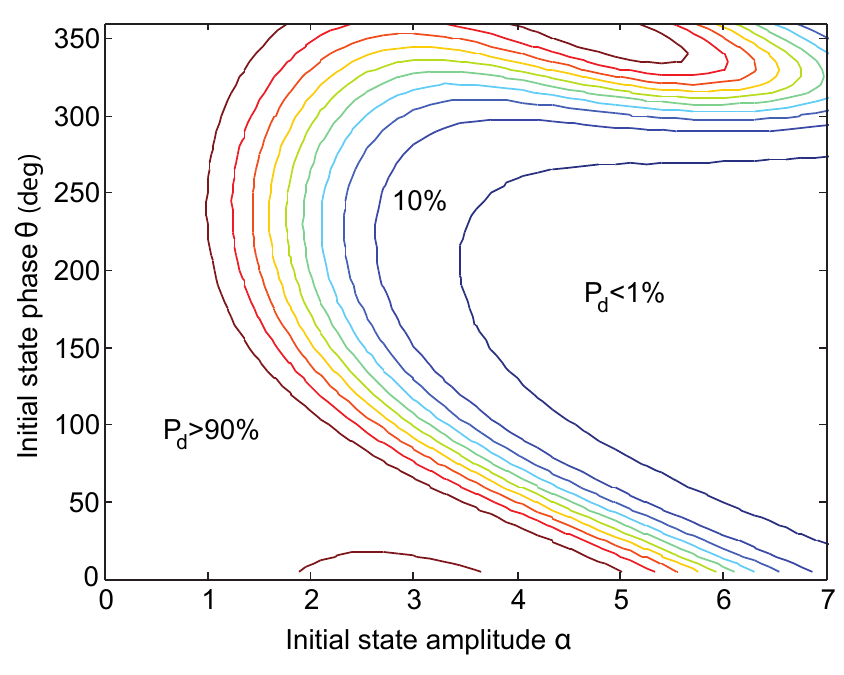}
\caption{\label{gino:basin_figure}Probability for the QCS to decay after being initialized as coherent state wave packet $|\alpha e^{i\theta}\ra$ and driven for a time $\kappa^{-1}$. The equal probability contours trace out two basins of attraction: states initialized inside the low decay probability contour ($P_d<1\%$) end up long lived ($\tau_b\gg \kappa^{-1}$), whereas states initialized left of $P_d=90\%$ quickly decay to the ground state and remain photon blockaded\index{photon blockade}. The parameters are $g/2\pi=100\textrm{MHz}$, $\kappa/2\pi=4.05\textrm{MHz}$, $\xi/2\pi=9.9\textrm{MHz}$,$(\om_{d}-\om_{c})/2\pi=12.3\textrm{MHz}$,$(\om_{c}-\om_{q})/2\pi=9.7\textrm{MHz}$, $T_1=1591ns$ ($\om_d$ is the drive frequency).}
\end{figure}

The QCS exist with drive amplitudes where the ground state is photon blockaded, giving
rise to a dynamical bistability\index{bistability} between quantum and semi-classical parts of the JC ladder. Indeed, we see that there is a basin of attraction for states initialized as coherent wave packets to persist as QCS, and we characterize it according to the probability of the state to decay on the timescale $\tau_b \gg \kappa^{-1}$. It is worth noting that this metastability cannot be described using a semiclassical picture\index{semiclassical} in cases where a significant part of the Hilbert space where the tail of the QCS resides is quantized (i.e. ``low lying" QCSs). Similarly processes with the opposite transition where the system succeeds in leaving of the blockade into the QCS basin cannot be described without solving the full quantum dynamics of the wave packet. The characterization and exact mechanisms of dynamical transitions between these states in this regime are a matter of ongoing research. Recent experiment have detected exponentially long lifetimes for quasi-coherent states \index{quasi-coherent states}excited by a chirp \index{frequency chirp} in a high excitation state of a Josephson phase qubit~\shortcite{katz_chirp} and driven by a holding tone. In Fig.~\ref{gino:basin_figure} we plot the contours of equal probability of the QCS to decay to a manifold of states close to the ground states, after a time $\kappa^{-1}$, given that it was initialized with a certain amplitude ($\alpha$) and phase ($\theta$). We see a large region supporting QCS, and the phase sensitivity can be understood qualitatively from the time dependent simulations: a mismatch between the phases of the drive and initial coherent state causes ringing of the wave packet outside of the basin into the too anharmonic part of the ladder from which it cannot recover. In addition to the existence of this basin, the anharmonicity acts together with the cavity decay to induce mixing of the QCS: even for bright states ($\bar{n}>40$) we extracted a relatively low purity of $\operatorname{Tr}(\rho^2)<0.5$.

\section{Applications for high fidelity qubit state measurement}\label{gino:applications}
\sectionmark{Applications to qubit state measurement}

The nonlinear response either in the dispersive or degenerate regimes opens the possibility of using the system's own high gain for (self) qubit readout. The figures of merit of a useful scheme involve low error rate, high contrast, speed and ideally the property of quantum non-demolition. To these we should probably add robustness because for a protocol to be useful it should sustain experimental imperfections. Fundamentally, the source of the high gain in the two regimes is in the nonlinearity of the two-level system (usually the two lowest levels of a quantized anharmonic oscillator). The meaning of nonlinearity and high gain for the semiclassical vs.\ quantum degenerate regimes are quite different and will be discussed below. What is common in these two schemes is that both require the application of relatively strong tones\index{strong!driving}, they both make use of a large part of the phase space of the system, and both are projective. In addition both schemes have many degrees of freedom in frequency and amplitude modulation of their control pulses and are therefore amenable to optimization.
For the scheme operating in the quantum degenerate regime\index{quantum degenerate regime} we have implemented a partial optimization in order to demonstrate that high fidelity\index{fidelity} readout is feasible.

\subsection{Symmetry breaking for the transmon device}\label{gino:sec:symbreak}

\begin{figure}[tb]
\centering
\includegraphics[scale=1.2]{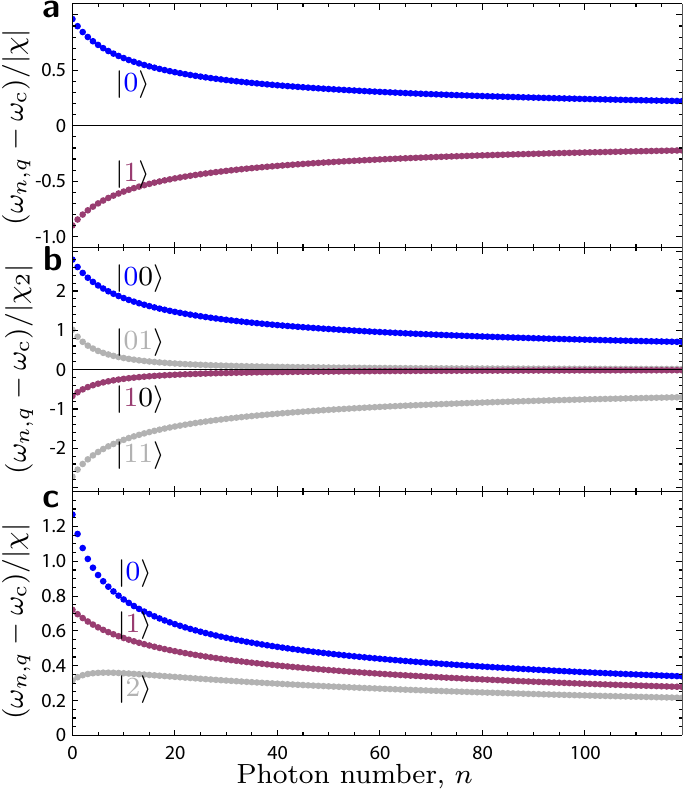}
\caption[XXX broken OUP style needs this because of shortcite in caption XXX]{(Color) Symmetry breaking. State-dependent transition frequency $\omega_{n,q}=2\pi(E_{n+1,q}-E_{n,q})$ versus photon number $n$, where $E_{n,q}$ denotes energy of the system eigenstate with $n$ photons and qubit state $q$:
(a)~for the JC model, parameters as in Figs.~\ref{gino:fig:latch000} and~\ref{gino:fig:densclass};
(b)~for the model extended to 2 qubits, $\delta_{1}/2\pi=-1.0\,\text{GHz}$, $\delta_2/2\pi=-2.0\,\text{GHz}$, $g_1/2\pi=g_2/2\pi=0.25\,\text{GHz}$. Here, $\chi_2$ denotes the 0-photon dispersive shift \index{dispersive regime}of the second qubit;
(c)~for the model extended to one transmon qubit~\shortcite{koch_charge-insensitive_2007}, tuned below the cavity, $\omc/2\pi=7\,\text{GHz}$ $E_C/2\pi=0.2\,\text{GHz}$, $E_J/2\pi=30\,\text{GHz}$, $g/2\pi=0.29\,\text{GHz}$. (For the given parameters, $\delta_{01}/2\pi=-0.5\,\text{GHz}$, $\delta_{12}/2\pi=-0.7\,\text{GHz}$, defining $\delta_{ij}=E_j-E_i-\omc$, with $E_i$ the energy of the $i$th transmon level.) In all panels, the transition frequency asymptotically returns to the bare cavity frequency. In (a) the frequencies within the $\sigma_z=\pm1$ manifolds are (nearly) symmetric with respect to the bare cavity frequency. For (b), if the state of one (`spectator') qubit is held constant, then the frequencies are asymmetric with respect to flipping the other (`active') qubit. In (c), the symmetry is also broken due the existence of higher levels in the weakly anharmonic transmon.}
\label{gino:fig:return}
\end{figure}

From the semiclassical eqns~\ref{gino:eq:classic}, \ref{gino:eq:chi} it follows that for $A\gg 1$ the response of the system will have an approximate symmetry of reflection with respect to the bare cavity frequency $A(\Omega,\sigma_z=+1)\approx A(-\Omega,\sigma_z=-1)$. Therefore the response at the bare cavity frequency will be nearly independent of the state of the qubit, with respect to both the low and high power regimes.
In order to translate the high gain available at the step into a qubit readout, it is necessary to break the symmetry of the response of the system between the qubit ground and excited states, such that the upper critical power $\xi_{C2}$ will be qubit state dependent.
In the JC model the symmetry follows from the weak dependence of the decoupled Hamiltonian $\tilde{H}$ on the qubit state for high photon occupation. However, the experimentally-observed  state dependence may be explained by a symmetry breaking caused by the higher levels of the weakly anharmonic transmon\index{transmon}, or by the presence of more than one qubit, see Fig.~\ref{gino:fig:return}. Such an asymmetric qubit dependent response is well known for the transmon in the {\em low power} dispersive regime \shortcite{koch_charge-insensitive_2007}\index{dispersive regime}. For the high power regime, such a dependence was shown in~\shortcite{maxime_reedout} by taking into account the full nonlinearity of the transmon.

When designing a readout scheme that employs such a diminishing anharmonicity, the contrast of the readout is a product of both the symmetry breaking and the characteristic nonlinear response of the system near the critical point $C_2$. Experiments~\shortcite{reedout_expt} initially were able to use this operating point to provide a scheme for qubit readout, which is attractive both because of the high fidelities achieved (approaching $90\%$, significantly better than is typical for linear dispersive readout in circuit QED~\shortcite{wallraff_approaching_2005,steffen_high_2010}) \index{Circuit QED}and because it does not require any auxiliary circuit elements in addition to the cavity and the qubit. This nonlinear response was also used for characterizing three-qubit GHZ states~\shortcite{leo_ghz_2010} and very recently for the characterization of transmon qubits\index{transmon} designed in a three dimensional architecture \shortcite{hanhee_2011}.

\subsection{Coherent control in the quantum degenerate regime}
\index{coherent contol}
Qubit readout in solid state systems is an open problem, which is currently the subject of intensive experimental and theoretical research.
High-fidelity \index{fidelity}single-shot readout is an important component for the successful implementation of quantum information protocols, such as measurement based error correction codes \shortcite{nielsen_chuang} as well as for closing the measurement loophole in Bell tests \shortcite{Garg_Mermin_Bell_1987,kofman_analysis_2008,ansmann_violation_2009}. For measurements where the observed pointer state depends linearly on the qubit state, for example dispersive readout in circuit QED (cQED)\index{Circuit QED} \shortcite{blais_cavity_2004}, there exists a unified theoretical understanding \shortcite{clerk_rmp_2009}. Experimentally, these schemes require a following amplifier \index{amplifier}of high gain and low noise, spurring the development of quantum limited amplifiers \shortcite{Lafe_amplifier,bergeal-2009,castellanos-beltran_amplification_2008}. However, the highest demonstrated fidelities to date rely on nonlinear measurement schemes with qubit dependent latching into a clearly distinguishable state, e.g. Josephson Bifurcation Amplifier (JBA)\index{bifurcation} as well as optimized readout of phase qubits \shortcite{siddiqi_dispersive_2006,mallet_single-shot_2009,ansmann_violation_2009}. For this class, during the measurement, the system evolves under the influence of time varying external fields and nonlinear dynamics, ultimately projecting the qubit state. The space for design and control parameters\index{coherent contol} is very large, and the dependence of the readout fidelity on them is highly nontrivial. Therefore the optimization is difficult and does not posses a generic structure.

\begin{figure}[tpb]
\includegraphics[scale=1.2]{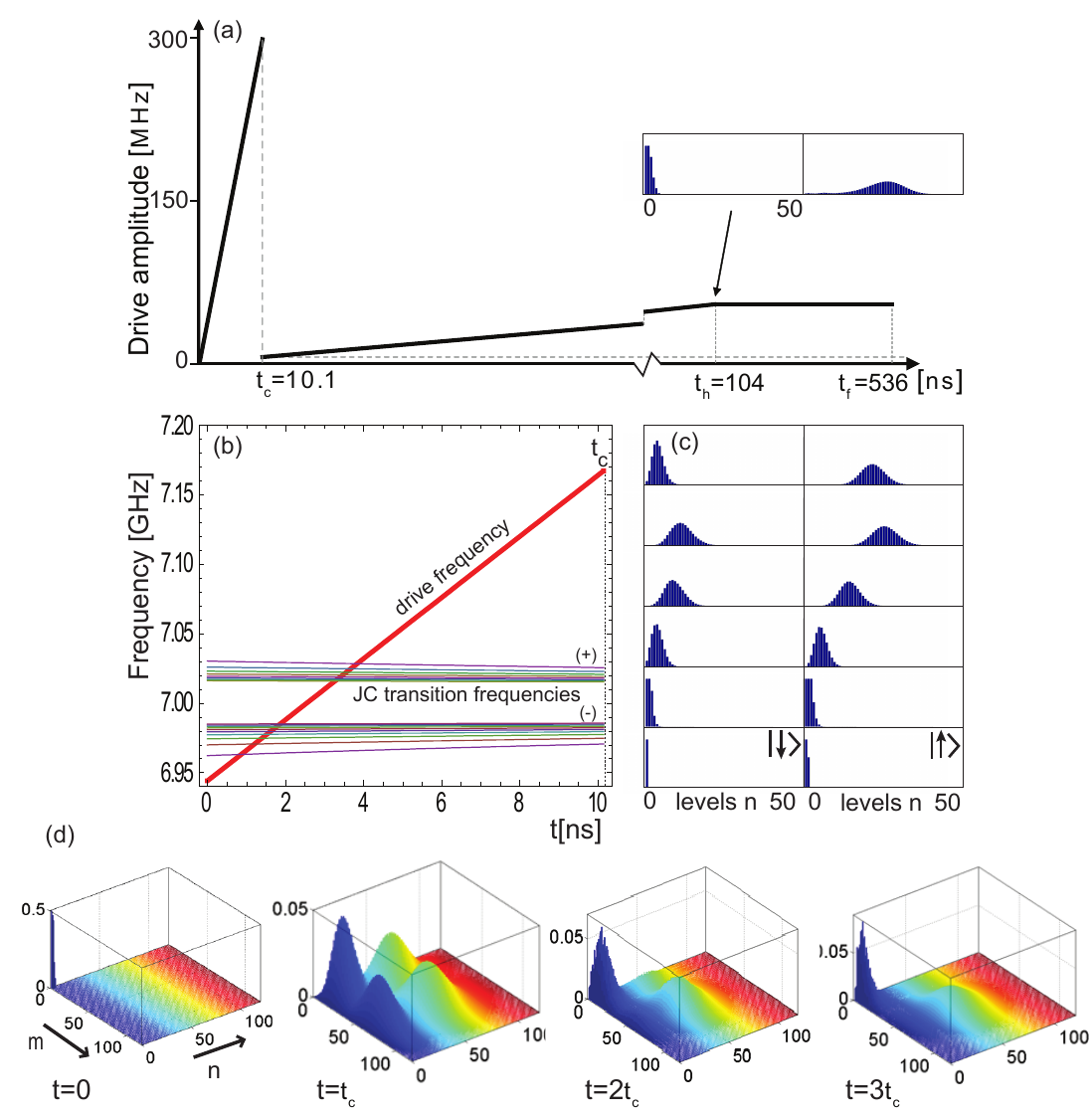}\centering
\caption{\label{gino:chirp_figure} (Color) Readout control pulse (a) Time trace of the drive amplitude: a fast initial chirp \index{frequency chirp}($10\,\text{ns}$) can selectively steer the initial state, while the qubit is detuned from the cavity ($(\om_q-\om_c)/2\pi \approx 2g$). It is followed by a slow displacement to increase contrast and lifetime of the latching state, while the qubit is resonant with the cavity ($\kappa/2\pi=2.5\,\textrm{MHz}$). The drive amplitude ramp is limited so that the photon blockade \index{photon blockade}is not broken, but the contrast is enhanced by additional driving at the highest drive amplitude. (b) A diagram of transition frequencies shows how the drive frequency chirps through the JC ladder frequencies of the (+) manifold, and how the manifold changes due to the time dependent qubit frequency. (c) Wave packet snapshots at selected times (indicated by bullet points on panel (b)) of the chirping drive frequency of panel (b) conditioned on the initial state of the qubit. (d) The temporal evolution of the reduced density matrix $|\rho_{mn}|$ (the $x,y$ axes denote the quantum numbers $m,n$ of the cavity levels) of the cavity with the control pulse (a) when the qubit initial state is superposition $\frac{1}{\sqrt{2}}(|0\ra+|1\ra)$. The resonator enters a mesoscopic state of superposition around $t=t_c$ due to the entanglement with the qubit and the quantum state sensitivity of the protocol. At later times the off-diagonal parts of this superposition dephase quickly due to the interaction with the environment and the state of the system is being completely projected around $t=3t_c$. }
\end{figure}

\begin{figure}[tb]
\includegraphics[width=\textwidth]{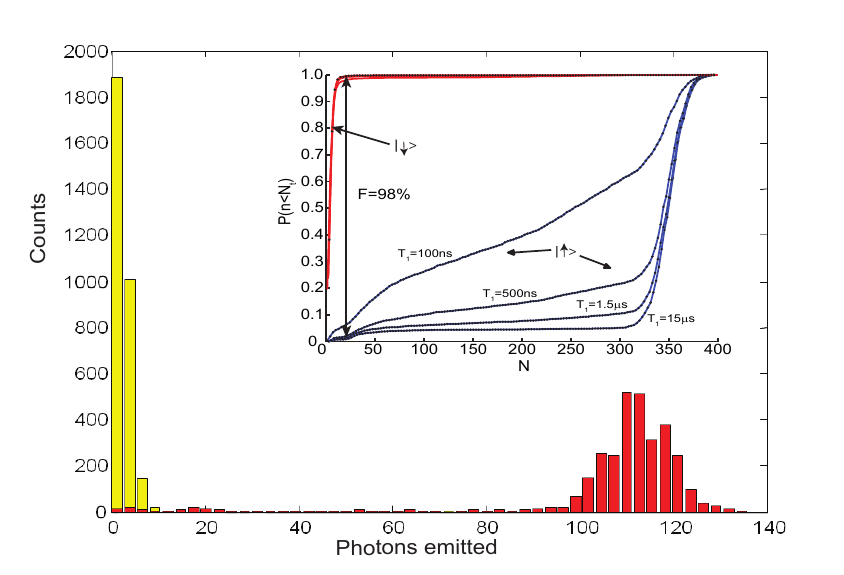}\centering
\caption{\label{gino:hist_figure} Example photon emission counts histogram generated by quantum trajectories simulations for the readout protocol of Fig.~\ref{gino:chirp_figure} with optimized parameters. Inset: For a longer holding tone, cumulative probability distributions for the number of photons (N) emitted from the cavity during the driving time $t_f$, for different qubit decay times ($T_1$), including a very long $T_1=15\mu s$ indicating that $T_1$ is not limiting the readout fidelity. For low detection thresholds ($N_{th}\approx 20$) for distinguishing $|\hspace{-0.05in}\uparrow\ra\,(N>N_{th})$ from $|\hspace{-0.05in}\downarrow\ra \,(N<N_{th})$ the fidelity can be very high ($>98\%$) for realistic values of qubit decay in cQED (a few microseconds), and of high contrast for more moderate fidelities ($>90\%$). The distributions also show a almost no false positives for higher thresholds $N_{th}>20$ (here $t_h=194ns$ and other parameters as for Fig.~\ref{gino:chirp_figure}).}
\end{figure}

We propose a coherent control based approach to the readout of a qubit that is strongly coupled to a cavity\index{strong!coupling}, based on an existing cQED architecture, but not necessarily limited to it. This approach is in the spirit of the latching readout schemes, but it differs in that the source of the nonlinearity is the Jaynes-Cummings (JC) interaction\index{Jaynes--Cummings}. When the qubit is brought into resonance with the cavity mode, the strong anharmonicity of the JC ladder of dressed states can prevent the excitation of the system even in the presence of a strong drive\index{strong!driving}, a quantum phenomenon known as photon blockade \index{photon blockade} \shortcite{imamolu_strongly_1997,birnbaum_photon_2005,houck-aps-2010}. However, due to fact that the JC anharmonicity is diminishing with the excitation number, we find a form of bistability\index{bistability}, where highly excited quasi-coherent states (QCS)\index{quasi-coherent states} co-exist with the blockaded dim states (Fig.~\ref{gino:basin_figure}).
In order to make use of this bistability to read out the qubit, it is necessary to solve the coherent control problem \index{coherent contol}of
selective population transfer, which is how to steer the system towards either the dim state or the QCS, depending on the initial state of the qubit (Fig.~\ref{gino:chirp_figure}). This selective dynamical mapping of the qubit state to the dim/bright states constitutes the readout scheme. It is potentially of high contrast, and hence robust against external amplifier \index{amplifier}imperfections.
An advantage of this readout is that it uses no additional components beyond the qubit and the cavity, both already present as part of the cQED architecture. Based on a full quantum simulation which includes dissipation of the qubit and the cavity (we ignore pure dephasing\footnote{Typically transmon qubits operate in a regime of $E_J/E_C\gg 1$ where pure dephasing is exponentially suppressed.}), we predict that implementing this scheme should yield very high fidelities between  $90\%$ and $98\%$ for a typical range of realistic cQED parameters (Fig.~\ref{gino:hist_figure}).


In the regime of coexistence, the dim quantum state and QCS present us with the possibility of implementing a high contrast readout scheme. This requires the solution of the coherent control problem\index{coherent contol} of steering the logical $|\hspace{-0.05in}\uparrow\ra$ state to some point on within the basin of attraction (Fig.~\ref{gino:basin_figure}), while keeping the $|\hspace{-0.05in}\downarrow\ra$ far from the basin, in the manifold of dim states. In the presence of dissipation, the latter would quickly decay to the ground state, and remain there even in the presence of driving, due to the photon-blockade, whereas
the QCS would persist for a long time $\tau_{b}$ and emit approximately $\kappa \la n \ra_{b} \tau_{b}$ photons.
The standard coherent control problem of population transfer \shortcite{bergmann_coherent_1998}, which was also discussed recently for superconducting qubits \index{superconductivity}\shortcite{Jirari-EPL}, is to maximize the probability $P_{i\rightarrow f}$ of steering the state $|i\ra$ to the state $|f\ra$. However, here the goal is to bring the probability for {\em selective} steering $P_{i\rightarrow f}+P_{i'\rightarrow f'}<2$ close to its theoretical maximum, which is an essentially different coherent control problem. For systems with very large anharmonicities, for example atomic systems it is possible to effectively implement a population transfer via adiabatic control \index{coherent contol}schemes such as STIRAP \shortcite{bergmann_coherent_1998}. The JC ladder anharmonicity is relatively small compared to atomic systems, and so these schemes are inapplicable here.

The control pulse sequence we apply is depicted in Fig.~\ref{gino:chirp_figure}, and consists of three parts: (1) a strong chirped pulse \index{frequency chirp}\index{strong!driving}($t<t_c$) drives the cavity and passes through the resonance of the cavity, with the qubit being detuned. Due to the interaction with the qubit, the cavity behaves as nonlinear oscillator with its set of transition frequencies depending on the state of the qubit (see the two distinct sets of lines in Fig.~\ref{gino:chirp_figure}(b)). The cavity responds with a ringing behavior which is different for the two cases (see Fig.~\ref{gino:chirp_figure}(c)).
The ringing due to the pulse effectively maps the $|\hspace{-0.05in}\downarrow\ra$ and $|\hspace{-0.05in}\uparrow\ra$ to the dim and bright state basins, respectively (see Fig. 3(c)). Since $\kappa t_c\ll 1$, an initial superposition $\alpha|\hspace{-0.05in}\uparrow\ra+\beta|\hspace{-0.05in}\downarrow\ra$ maps into a coherent superposition of the dim and bright states. Next, (2) a much weaker long pulse transfers the initially created bright state (for initial $|\hspace{-0.05in}\uparrow\ra$) to even brighter and longer lived states ($t_c<t<t_h$), and (3) steady driving for additional contrast ($t_h<t<t_f$). For an initial superposition the interference terms between the dim and bright states decohere on the timescale of $\kappa^{-1}$, such that the interaction with the reservoir for $t>t_c$ effects a projection of the pointer state. In designing such a pulse sequence we have the following physical considerations: (a) the initial fast selective chirp\index{frequency chirp} has to be optimally matched to the level structure so that the population transfer and selectivity would be extremely high (b) it is necessary to chirp up quickly before decay processes become effective and result in false negative counts ($t_c\kappa\approx 0.16$) (c) for $t>t_c$ it is necessary to drastically reduce drive strength, since it reaches drive strengths which would break the photon blockade through multiphoton processes if it persisted. The piecewise linear chirp sequence is fed into a full quantum simulation that includes decay, and the 13 parameters of the system and drive are optimized with respect to the total readout fidelity.\index{fidelity}

The cumulative probability distributions to emit $N$ photons conditioned on starting in two initial qubit states are plotted in Fig.~\ref{gino:hist_figure}. These distributions were optimized for $T_1=1\mu s$ and then regenerated after varying $T_1$, in order to depict the effect of qubit relaxation on the readout. There are two figures of merit from figure~\ref{gino:hist_figure}: one is that there exist very high fidelities $\mathcal{F}=1-P({\uparrow|\downarrow})-P({\downarrow|\uparrow})$ exceeding $98\%$ for a low threshold around $N_{th}=20$ even for relatively short lived qubits ($T_1\approx 500\,\text{ns}$). In order to take advantage of these fidelities a very low noise amplifier would be needed.\index{amplifier}
In addition, we find high contrast and high fidelities ($>90\%$) for long lived qubits $T_1>1.5\,\text{$\mu$s}$ with thresholds around $N_{th}=150$, which should be accessible with state-of-the-art HEMT amplifiers. The limit of obtainable fidelities with this control scheme \index{coherent contol}is not due to finite qubit lifetime, as we see from the curve that was simulated for $T_1=15\,\text{$\mu$s}$. The reason is that after the QCS is generated at $t=t_c$ the qubit is brought into resonance to form the blockade, and that enhances the anharmonicity of the system. Since the drive is not in resonance with all the transition frequencies relevant to the wave packet, this leads to the few percent of decay events which are unrecovered by the drive.
Note also that a useful feature of these distributions is the very low level of false positives (red curve for qubit state $|\hspace{-0.05in}\downarrow\ra$), for a wide range of thresholds, originating from the effectiveness of the photon blockade\index{photon blockade}.

\subsection{Experimental considerations}

For experimental applications it is useful to know how robust the fidelity\index{fidelity} is against deviations of the control pulse parameters from their optimal values. We therefore varied the parameters of the initial chirp\index{frequency chirp} pulse $\delta_{d,c}=\om_d-\om_c,\dot{\delta}_{d,c}, \delta_{d,q}=\om_d-\om_q, \dot{\delta}_{d,q}, \dot{\xi}, t_c,$ and $\kappa$ independently around their optimal values. Table \ref{gino:sens_tab} shows for each parameter the range of variation for which the fidelity is above $98\%$ (the range cited is the smaller of the two ranges above and below the optimal value).
\begin{table}\label{gino:sens_tab}\centering
\begin{tabular}{|c|c|c|}
\hline
Parameter & Optimal value & Range for $\mathcal{F}\ge 98\%$ \\
\hline
$\delta_{d,c}(t=0)/2\pi$    & -56.0 MHz     & $\pm 40\%$  \\
$\dot{\delta}_{d,c}/2\pi$   & 21.9 MHz/ns    & $\pm 20\%$  \\
$\delta_{d,q}(t=0)/2\pi$    & -226.2 MHz     & $\pm 30\%$  \\
$\dot{\delta}_{d,q}/2\pi$   & -8.2 MHz/ns   & $\pm 100\%$ \\
$\dot{\xi}/2\pi$            & 29.4 MHz/ns    & $\pm 60\%$  \\
$t_c$                       & 10.1 ns       & $\pm 10\%$  \\
$\kappa/2\pi$               & 2.5 MHz        & $\pm 60\%$  \\
\hline
\end{tabular}
\caption{Values of optimal chirp parameters for achieving maximal fidelity and their relative tolerances (given in percents of the optimal value).}
\end{table}
The fidelity is most sensitive to variations of the duration of the chirp\index{frequency chirp} pulse $t_c$ which yields a tolerance of $\pm 10\%$ and higher ranges ($>\pm 20\%$) for the rest of the parameters. For achieving slightly less high fidelities ($>97\%$) the bounds for $t_c$ increase significantly to $\pm 20\%$, which is important for a realistic experimental setup, since the quench of the Hamiltonian parameters at $t=t_c$ will take a few nanoseconds with the current microwave technologies.

The effect of amplifier noise should also be considered. A cryogenic HEMT amplifier with noise temperature of $T_N\approx 5K$ adds noise to the amplitude quadratures $b_{x}(z,t)=\frac{1}{\sqrt{2}}(b^{\dagger}(z,t)+ b(z,t))$, $b_{y}(z,t)=\frac{i}{\sqrt{2}}(b^{\dagger}(z,t)-b(z,t))$ of the input signal, where $b(z,t)$ denotes the annihilation of a photon in the transmission line at position $z$ and time $t$ \shortcite{clerk_rmp_2009}. The annihilation and creation operators are defined such that $\phi(z_0,t)=\la b^{\dagger}(z_0,t)b(z_0,t)\ra$ is the photon flux at the point of entry $z_0$ to the amplifier. The dimensionless spectral density of the noise is given by $S=k_B T_N/\hbar \omega \approx 20$ where $\omega$ is the frequency of the probe signal. To obtain an optimal signal to noise ratio (SNR) \index{signal to noise ratio}the signal is measured by time integration, which introduces a bandwidth of $1/t_f$ around the carrier frequency, where $t_f$ is defined in Fig. \ref{gino:chirp_figure}. This bandwidth is optimal \shortcite{gambetta_protocols_2007} in the sense that it lets in all the signal but keeps the noise level minimal (disregarding the short initial transient of the chirp\index{frequency chirp}). In this model the quadrature noise is assumed to be normally distributed with $b_{i,noise}(z,t)\sim\mathcal{N}(\mu=0, \sigma=\sqrt{S/t_f})$ where $i=x,y$. It is therefore possible to estimate the effect of the amplifier noise \index{amplifier}by analyzing the distribution of the total number of emitted photons, including the photons of noise, $N=\frac{1}{2t_f}\sum_{i=x,y}\left[\int_0^{t_f}(b_i(z_0,t)+b_{i,noise}(z_0,t))dt\right]^2$, where the noise is added to each quadrature independently. For the cases depicted in Fig.~\ref{gino:hist_figure} where $N_{\uparrow}\approx 350$ we have $SNR\approx 17$ and approximately $3\%$ of errors were added by the noise, which is therefore not a significant limit to fidelities in the $90-95\%$ range. For $T_N>5K$ the effect of the noise can be mitigated by increasing $t_f$ (up to time $\tau_b$).

\subsection{Comparison with other schemes}

For cQED all the necessary components for the above scheme have been experimentally demonstrated. Strong qubit-cavity coupling has been demonstrated in many experiments \shortcite{thompson_observation_1992,raimond_colloquium:_2001,wallra_strong_2004}. Strong driving of a cavity-qubit system has been shown in \shortcite{baur_measurement_2009}, with the system behaving in a predictable way, as well as photon blockade \shortcite{bishop_nonlinear_2009} and fast dynamical control \index{coherent contol}of the qubit frequency via flux bias lines \shortcite{dicarlo_demonstration_2009}. In addition there is evidence both theoretically and experimentally for the increasing role that quantum coherent control plays in the optimization of these systems for tasks of quantum information processing. As examples we can mention improving single qubit gates \shortcite{motzoi_simple_2009}, two-qubit gates \shortcite{fisher_optimal_2009}, and population transfer for phase qubits \shortcite{Jirari-EPL}. We therefore believe that the readout scheme would be applicable for the transmon\index{transmon}, although the control parameters would have to be re-optimized due to the effect of additional levels.

The suggested readout scheme is different from other existing schemes in several aspects. Compared to dispersive readout \shortcite{blais_cavity_2004} it involves very nonlinear dynamics and could potentially exhibit much higher fidelity \index{fidelity}and contrast. Even though it relies on a dynamical bistability\index{bistability}, it is essentially different from the JBA and the scheme from Section~\ref{gino:sec:symbreak} (see also \shortcite{bishop_jco}), since it explicitly operates using the quantum photon blockade\index{photon blockade}. Our scheme is also essentially different from a recently suggested adaptation of electron-shelving readout to circuit QED\index{Circuit QED} \shortcite{englert_mesoscopic_2010}. The latter makes use of a {\em third level} in addition to the two levels which define the qubit, requires a direct coupling to the qubit with negligible direct driving of the resonator, and strong driving of the qubit in the regime where the rotating wave approximation breaks down. It is important to stress that the optimization of the control parameters in our scheme is only partial, since we have limited ourselves to simple linear chirps in this work\index{frequency chirp}. Our optimization scheme involves a Simplex algorithm\index{Simplex} that searches for a local maximum for the readout fidelity. Each step the fidelity is calculated by solving the stochastic Schr\"{o}dinger equation \index{stochastic Schr\"{o}dinger equation}for the dynamical evolution for different realizations of the measurement signal. More complex modulations (e.g.~using GRAPE type algorithms) are certainly possible although the standard methods for optimal control \shortcite{khaneja_optimal_2005} may be difficult to implement here due to the large Hilbert space. Therefore we believe that an experimentally based optimization using adaptive feedback control \shortcite{judson_teaching_1992} might be best option, and has the potential to yield superior readout fidelities for higher detection thresholds.

\section{Conclusion and future prospects}

In the solid state realization of superconducting qubits\index{superconductivity}, nonlinear oscillator dynamics arise naturally from the quantum circuit Hamiltonians, operating in the semiclassical\index{semiclassical} or fully quantum regime. Some of the physical phenomena are associated with the effect of quantum noise on the oscillator state whereas some are associated with the quantized anharmonic ladder of states. In this review we tried to demonstrate the theoretical challenges that arise in the high excitation regime of these models and its relevance to state-of-the-art experiments.

Since the experimental methods are evolving rapidly, it becomes intriguing to develop theoretical schemes of control \index{coherent contol}which utilize larger parts of the accessible Hilbert space. This involves extending our understanding of nonlinear response of multiple strongly coupled transmon-cavity systems to the high drive power regime. We use analytical tools and exact simulations to access these regimes and test new protocols. It would also be increasingly important to adapt tried and tested optimal control techniques to these systems as a support of existing and future experiments.

This work was supported by the NSF under Grants Nos.~DMR-1004406 and DMR-0653377, LPS/NSA under ARO Contract No.~W911NF-09-1-0514, and in part by the facilities and staff of the Yale University Faculty of Arts and Sciences High Performance Computing Center.

\bibliography{levisordinary,levapsjour,review}
\end{document}